\documentclass[twocolumn]{svjour3}
\usepackage{graphics,graphicx,psfrag,epsfig}
\journalname{Granular Matter}
\begin{document}

\title{Jet-induced 2-D crater formation with horizontal symmetry breaking}

\author{Abram H. Clark \and  Robert P. Behringer  }

\authorrunning{A. H. Clark, R. P. Behringer} 

\institute{A. Clark \and R. Behringer$^{*}$ \at
              Department of Physics, Duke University\\
              Tel.$^{*}$: +001 919 660-2550\\
              Fax$^{*}$: +001 919 660-2525\\
              \email{$^{*}$bob@phy.duke.edu}           
}

\date{Received: date / Accepted: date}

\maketitle

\begin{abstract}
 
We investigate the formation of a crater in a 2-D bed of granular material by a jet of impinging gas, motivated by the problem of a retrograde rocket landing on a planetary surface. The crater is characterized in terms of depth and shape as it evolves, as well as by the horizontal position of the bottom of the crater. The crater tends to grow logarithmically in time, a result which is common in related experiments. We also observe a horizontal symmetry breaking at certain well-defined conditions which, as we will demonstrate, could be of considerable practical concern for lunar or planetary landers. We present data on the evolution of these asymmetric states and attempt to give insights into the mechanism behind the symmetry-breaking bifurcation.

\end{abstract}

\section{Introduction}

\begin{figure*}[h!tbp]
\centering
\includegraphics[trim=10mm 0mm 10mm 0mm, width=\textwidth]{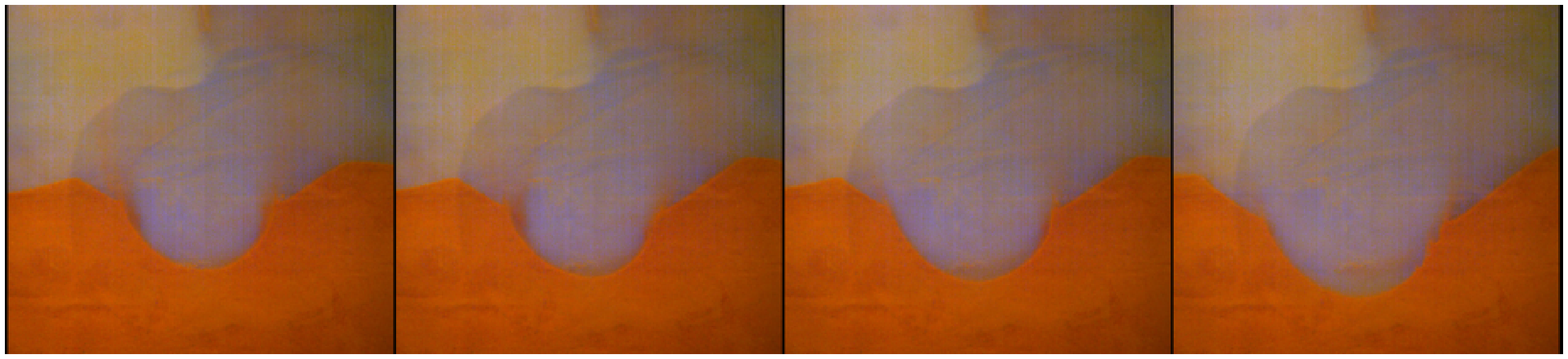}
\includegraphics[trim=8mm 0mm 8mm 15mm, width=\textwidth]{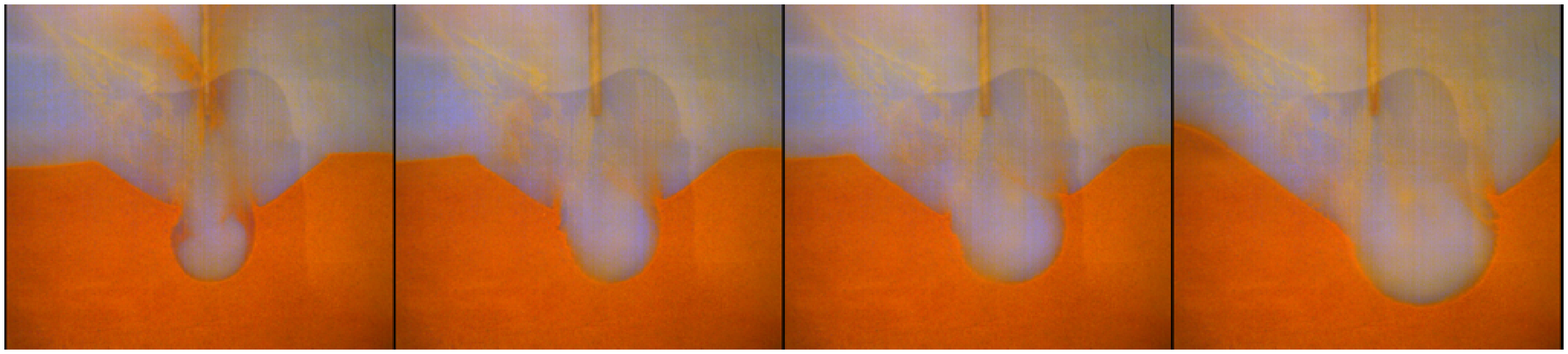}
\caption{Still frames of a weak, leftward (TOP) and a strong, rightward (BOTTOM) symmetry breaking process (for both sets of frames: left to right, at 3, 5, 7, and 15 seconds). Weak symmetry breaking typically occurs with higher jet heights and lower pressures, while strong symmetry breaking typically occurs with lower jets and higher pressures. The weak case shown has a jet starting height of $h=75$~mm and pressure of $P=7$~psi (see text for details of experimental setup). Throughout the process, the crater retains its shape fairly well and debris is ejected from both sides of the crater. For the strong case shown, with $h=35$~mm and $P=10$~psi, the debris is ejected more vertically, leading to a strong coupling between the two sides of the crater. As the crater shifts to the right, debris is ejected only from the right side of the crater, and a circular flow pattern emerges (counterclockwise, in images shown here).}
\label{symbreakstrong}
\end{figure*}

Recent initiatives for expanding human presence in the solar system, including further exploration of the Moon and Mars, have generated interest in understanding planetary crater formation from a retrograde rocket \cite{metzger1,metzger2,metzger4,metzger5,metzger3,Kuang}. This problem essentially amounts to a jet of gas impinging on a granular bed from above. In addition to its practical application, this fundamental two-phase flow (gas and granular material) comprises an interesting problem in it's own right. Depending on the strength and composition of the jet and the characteristics of the granular material, the crater formation can occur in different ways (e.g. surface erosion of particles, penetration of the gas into the granular material, bulk shear of the granular bed) \cite{metzger1}. However, we will restrict ourselves primarily to considering so-called viscous erosion \cite{Kok}, in which the jet lifts or rolls the top few layers of grains, and a scour hole forms beneath the jet. This is the easiest regime to reach in that it takes the smallest jet velocity to accomplish. Additionally, due to a pulse when the jet is switched on, the initial formation of the craters in our experiments contains a component of eruption from gas diffusing into the interstices. The strength of this effect depends strongly on the velocity of the jet. In our experiments, we used a high-speed camera to observe the crater formation in a quasi-2D environment,  characterizing the process by the time-evolution of the shape, size, and location of the crater, as we varied control parameters of the system. During this process, we observed an unexpected symmetry breaking: under certain conditions, the centralized crater becomes unstable, and the crater moves to the left or to the right. Figure~\ref{symbreakstrong} shows frames from videos, where symmetric craters break symmetry in two different modes which we denote ``strong'' and ``weak'', which will be the focus of the bulk of this paper. We will attempt to give a coherent picture of the cratering process as a whole, including the symmetry-breaking effect, which is consistent with the data and gives physical insights into the process, at least on a phenomenological level.

\section{Experimental Setup}

\begin{figure}[h!tbp]
\centering
\includegraphics[width=0.9\columnwidth]{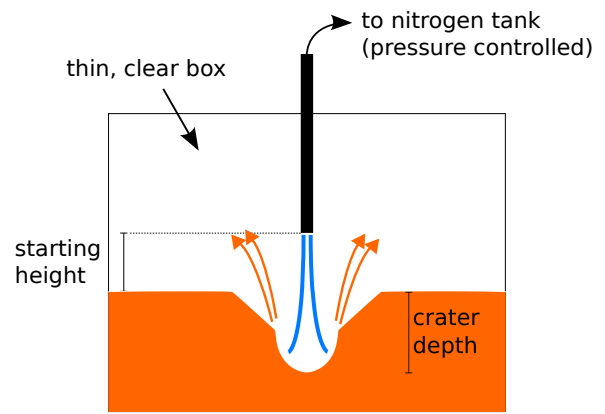}
\caption{Sketch of the experimental apparatus}
\label{SetupPic}
\end{figure}

The primary apparatus used in our experiments (Fig. \ref{SetupPic}) consists of a thin, transparent acrylic box 
with inner dimensions approximately $25\times 50\times 1$ cm. The box was partially 
filled with various granular materials, each of which possesses different characteristics which clearly 
affected the dynamics of the cratering process, including an earth-source red sand, as well as lunar 
and Martian soil simulants manufactured by Orbital Technologies (ORBITEC). A thin metal pipe (22 cm long, with inner diameter of 
0.44 cm) was inserted through a hole in the top of the box. The pipe was connected to a pressure-controlled 
tank of nitrogen gas through a flexible tube: 40 cm with inner diameter 0.89 cm, then a valve, then 37 cm with inner diameter 0.38 cm to the pressure sensor and nitrogen tank. The total length of the path between the pressure sensor and the nozzle is $\sim 1$~m. We varied $P$, the pressure driving the gas flow between 2 and 10 psi, $h$, the starting height of the jet above the granular material between 35~mm and 75~mm, and the type of material used, i.e., the three materials discussed above. Using high-speed videos at 100 frames per second, we characterized the cratering process by measuring $D$, the depth of the crater from the initial surface as a function of time, $w$, the width and general shape of the crater as it evolved, and $A_x$, the horizontal displacement of the bottom of the crater with respect to center, which served as a measurement of the horizontal asymmetry.

Note that this apparatus is different from the ones used in previous experiments by Metzger et al. \cite{metzger1,metzger2,metzger4,metzger5}, in that it is quasi-two-dimensional instead of a three-dimensional half-space (where the jet driving the crater formation is placed against a transparent side wall of a container, effectively taking a cross-section of a 3D cratering experiment). These two approaches are both chosen for their obvious advantages over the fully 3D case in terms of taking measurements. By making either simplification, one obviously changes the system. For the primary crater formation, the quasi-2D system exhibits behavior which is very similar to the 3D half-space.

The material properties were as follows: the sand consisted of roughly monodisperse (about 1 mm diameter) but rough particles with moderate macroscopic friction (internal angle of friction, $\delta=37$ degrees), the lunar simulant, JSC-1, was very frictional ($\delta=45$ degrees) and had a broad distribution of particle sizes (0.01 - 2 mm) (see \cite{mckay,metzger5}). The 
Mars simulant, JSC Mars-1, was less frictional then the lunar simulant ($\delta=36$ degrees) and also had a broad size distribution 
(0.02 - 1 mm) \cite{allen}. 

\section{Experimental Results}

\subsection{Symmetric Crater Evolution}

A general qualitative description of the process by which cratering occurred is as follows: the impinging jet of gas created an elevated pressure and an upward flow at the bottom of the crater. The flow entrained a relatively thin layer of granular material along the sides and bottom of the crater, and drove this material out of the crater. As one might expect, the smaller, lighter particles tended to become entrained in the gas flow much more easily than heavier particles. Hence, in materials with a large distribution of sizes, such as the lunar or Martian simulants, the flow of debris exiting the crater typically contained a higher concentration of smaller particles. We note that the regime of gas flow in our experiments was one where the gas had enough kinetic energy to move the particles, but not enough kinetic energy to penetrate the granular material in a way that caused eruption. 

We consider first what is probably the most obvious measurement in a cratering process, namely, the depth of the crater, $D(t)$. In all the data, 
the depth of the crater was found to grow logarithmically with time during the initial stages:

\begin{equation}
D(t) =a \ln{(b t +1)}
\label{eq:depthlog}
\end{equation}


This basic form governs the depth evolution for a substantial period of time after the jet is turned on, and it is applicable across a wide range of parameter values for jet speed, jet starting height, and material used, as long as the digging process is one of viscous erosion. The logarithmic growth in these quasi-2D experiments is similar to the observations by Metzger et al. \cite{metzger1} in a 3D half-space. Logarithmic growth is observed in cratering studies with widely varying experimental conditions \cite{metzger1,metzger2,rajaratnam,mazurek,varas,kellay}, which suggests an underlying physical principle which manifests itself irrespective of the details. 

\begin{figure}[h!tb]
\centering
\includegraphics[width=\columnwidth]{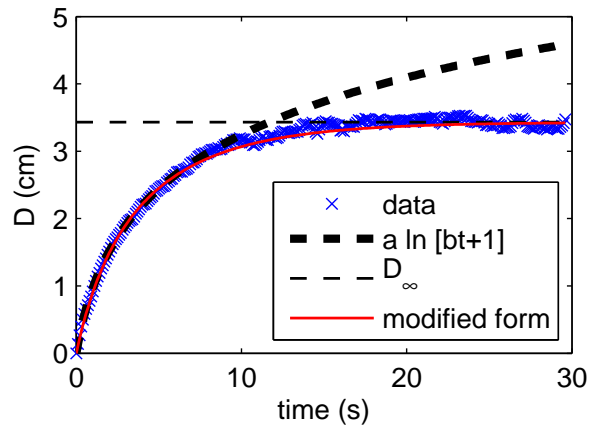}
\caption{Above is data demonstrating the asymptotic crater depth. The crater grows logarithmically in time 
for the first 10 seconds (top). However, observing for a longer time, we clearly see the depth approach an asymptote, $D_\infty$ (bottom). This data is from a cratering experiment with red sand in which there was no breaking of the horizontal symmetry, as will be discussed later. We later discuss the dashed red line below in the context of a growth model which allows for saturation.}
\label{saturation}
\end{figure}

The formation of the crater was not simply determined by ejection of material, but rather, involved competition between digging and refilling processes. As the crater grew deeper and wider, ejected debris had a greater chance of being recirculated back into the crater, by either falling directly in the crater or by avalanching in from an outer crater which formed in some cases and is discussed below. After a period of logarithmic growth, the crater eventually approached an asymptotic depth, $D_{\infty}$,
where the ejection of matter by driving gas is balanced by a reflux of grains back into the crater. In these experiments, this saturation point seemed to depend on the box geometry, including its size. We return later to this issue, where we suggest an evolution equation which captures the initial logarithmic growth and the asymptotic depth. Characteristic data is shown in Fig. \ref{saturation}.

Eq.~\ref{eq:depthlog} requires a characteristic length scale ($a$), and a characteristic time scale ($1/b$). An important issue concerns how the dependence of $a$ and $b$ depend on other system parameters. In a 3D half-space experiment \cite{metzger2}, Metzger, et al, investigated the dependence of these parameters on the starting height of the nozzle 
($h$), the kinetic energy density or momentum flux of the gas ($\rho v^2$), and the inner diameter of the nozzle ($d$). These authors found that $a$ scales linearly with $h$ and is unaffected by $\rho v^2$ or $d$, while $b$ scales linearly with $\rho v^2$, $h^{-3}$, and $d$. Our measurements agree with this: if we assume that the jet velocity scales linearly with the driving gas pressure, $P$, implying that the flow in the connecting line is close to laminar, our results are consistent with the Metzger et al. measurements.

\begin{figure}[h!tbp]
\centering
\includegraphics[width=0.85\columnwidth]{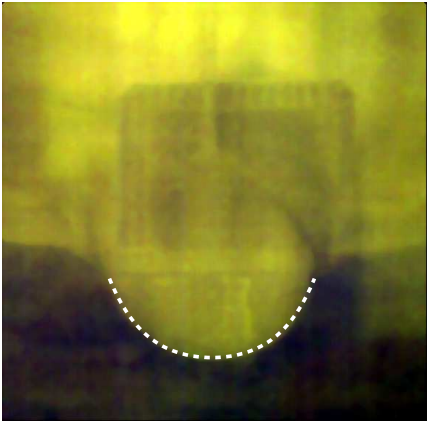}
\includegraphics[trim=0mm 0mm 0mm 5mm, width=0.85\columnwidth]{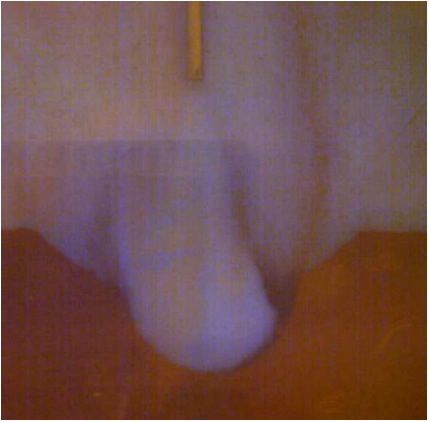}
\caption{Images taken during crater formation using lunar simulant (TOP) and Mars simulant (BOTTOM). For the lunar simulant, the crater shape is approximately a hyperbolic cosine (dashed line). The shape remains very consistent as the crater grows: depth and width both grow logarithmically in time with extremely similar time scales. For the Mars simulant, as well as for the sand used for the bulk of this paper (shown in Fig.~\ref{symbreakstrong}), reduced friction gives rise to inner and outer crater.}
\label{lunarcrater}
\end{figure}

The basic shape of the crater was most clearly affected by the frictional properties of the granular material, as in \cite{metzger5}. Less frictional materials, such as the sand and Mars simulant, Figs.~\ref{lunarcrater} (bottom) and Fig.~\ref{symbreakstrong}, formed craters with two clear sections: an inner crater, which is directly below the impinging jet and had rounded shape (discussed below), and the outer crater, which connected the outer edge of the inner crater along planes to the surface at approximately the angle of repose of the granular material. Highly frictional materials, i.e. the lunar simulant, did not exhibit a significant outer crater, at least not for time scales of these experiments. It seems likely that the outer crater is formed due to Coulomb failure when the slope of the side wall becomes steep. The more frictional materials can support a nearly vertical side wall with some added help from the pressure of the jet (see Figure \ref{lunarcrater}), whereas the less frictional materials cannot. 



\begin{figure}[ht]
\centering
\includegraphics[scale=0.9]{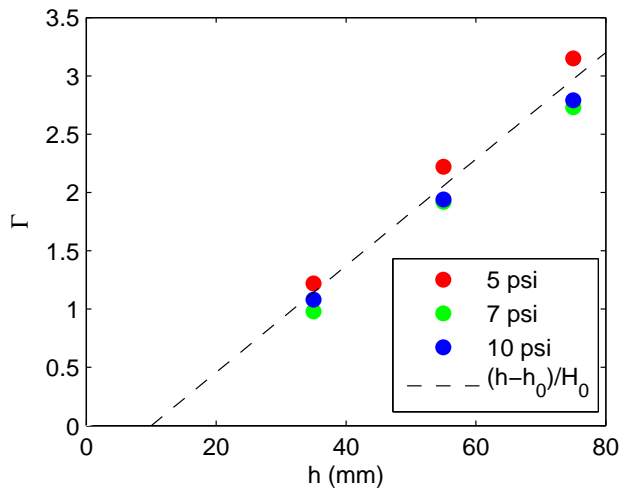}
\caption{Measurements of the aspect ratio for craters in red sand after 1 second: the jet height sets the crater's shape. The aspect ratio, $\Gamma$ which is measured as the width-to-depth ratio of the inner crater, is linearly dependent on jet height and only very weakly dependent on driving pressure.}
\label{aspectratio}
\end{figure}

Also of interest is the shape of the inner crater, which we characterize by the aspect ratio $\Gamma$, defined as the width-to-depth ratio of the inner crater.
As shown in Fig.~\ref{aspectratio}, the starting height of the jet has a strong influence on the fundamental geometry of the crater. In fact, Fig.~\ref{aspectratio} shows that $\Gamma$ scales linearly with the jet starting height, with at best a weak dependence on air speed/pressure. This dependence of $\Gamma$ on jet height can be seen in the top-left and bottom-left images of Fig.~\ref{symbreakstrong}: the top case has $h=75$~mm and $\Gamma\approx 2$, and the bottom case has $h=35$~mm and $\Gamma\approx 1$, consistent with Fig.~\ref{aspectratio}. We can approximate $\Gamma$ with the following form:
\begin{equation}
\Gamma = (h - h_0)/H_0,
\label{geom}
\end{equation}
where $h_0$ is about twice the tube inner diameter, and $H_0 \simeq 2.2$. 

\subsection{Horizontal Symmetry Breaking}

As previously stated, for certain conditions, the cratering process became horizontally asymmetric, with the crater moving to one side or the other. We observe two qualitatively different modes of horizontal symmetry breaking in the cratering process, shown in Fig.~\ref{symbreakstrong}: the top shows weak symmetry breaking, which is characterized by broad craters and slow transitions, and the bottom shows strong symmetry breaking, which is characterized by narrower craters and rapid transitions. This phenomenon was observed in the red sand and in the Martian simulant, but not in the Lunar simulant, even for very strong jets. This difference was presumably due to the frictional properties of the materials (that is, symmetry breaking requires a less frictional material). The following results concerning symmetry breaking are all from experiments using the red sand.

\begin{figure}[h!]
\centering
\includegraphics[trim=0mm 0mm 0mm 0mm, width=0.9\columnwidth] {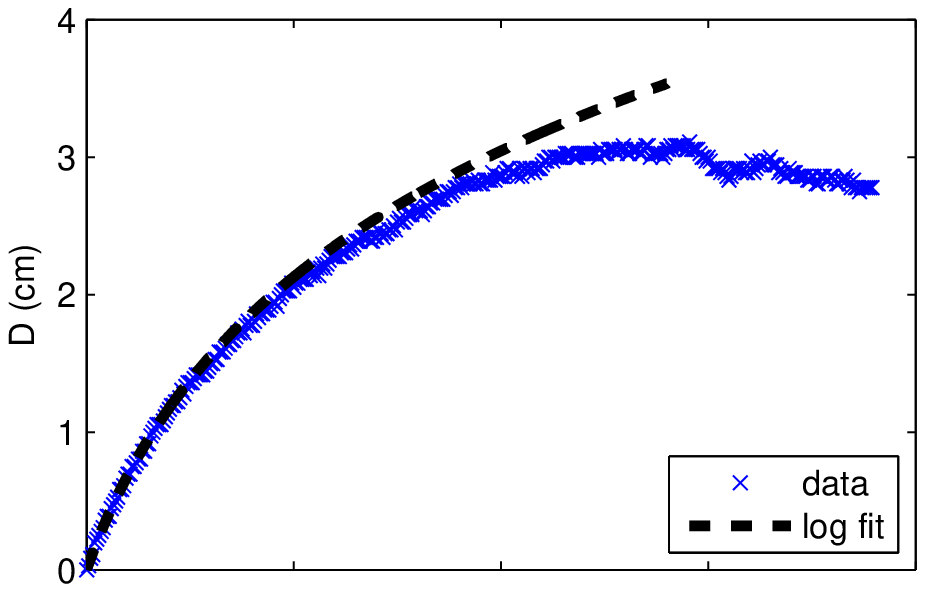}
\includegraphics[trim=1mm 0mm 0mm 6mm, width=0.9\columnwidth] {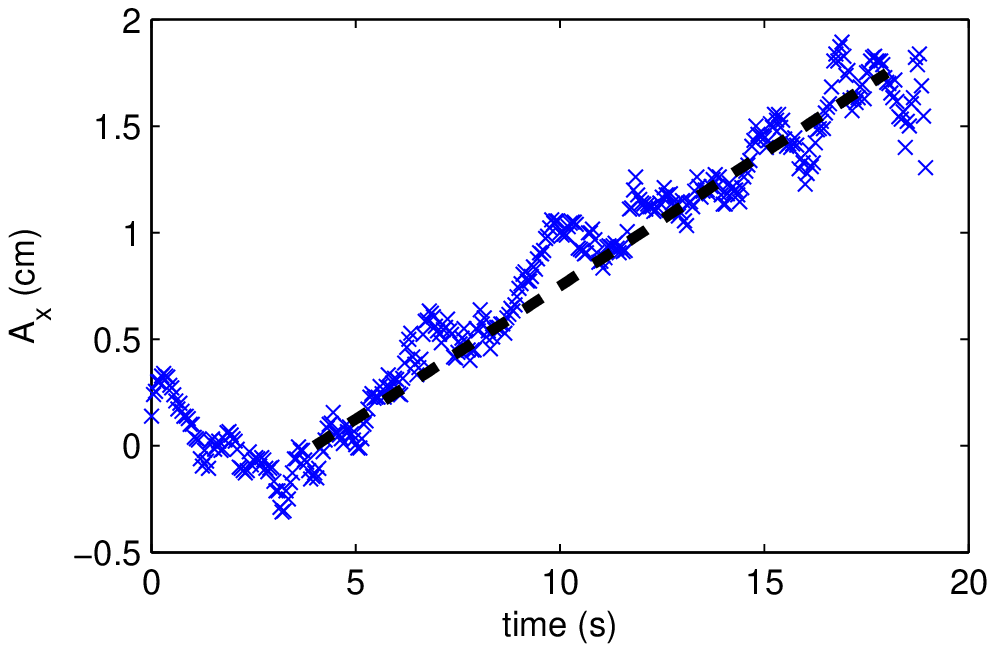}
\caption{Symmetry breaking, with $P=6$~psi and $h=35$~mm, for weak symmetry breaking, with depth (top) and asymmetry (bottom) plotted: points are experimental data, dotted lines show logarithmic and linear fits respectively, and the horizontal asymmetry, $A_x$, is the deviation from the center position. The lowest point of the crater moves quasi-linearly as it drifts from symmetric to asymmetric. At the end of the linear transition, the horizontal position asymptotes to its fully asymmetric state, where one edge of the inner crater is aligned with the jet.}
\label{weakdata}
\end{figure}

To quantify the symmetry breaking, we measure the depth, $D$, and horizontal deviation from the center, $A_x$, of the lowest point of the crater. Figure~\ref{weakdata} shows these quantities for a weak symmetry-breaking case. The logarithmic growth of depth is not substantially disrupted by the symmetry breaking, and at the same time the horizontal position of the crater changes gradually, i.e. quasi-linearly. 

\begin{figure}[th]
\centering
\includegraphics[trim=0mm 0mm 0mm 0mm, width=0.9\columnwidth] {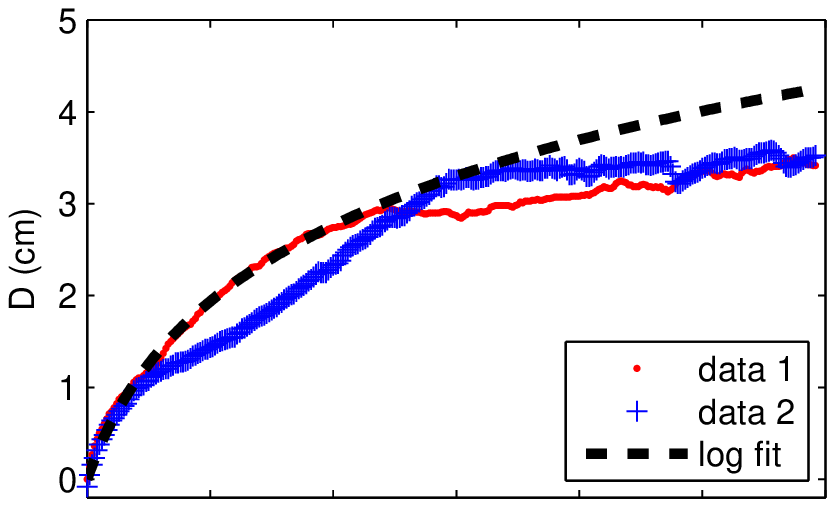}
\includegraphics[trim=0mm 0mm 0mm 5mm, width=0.9\columnwidth] {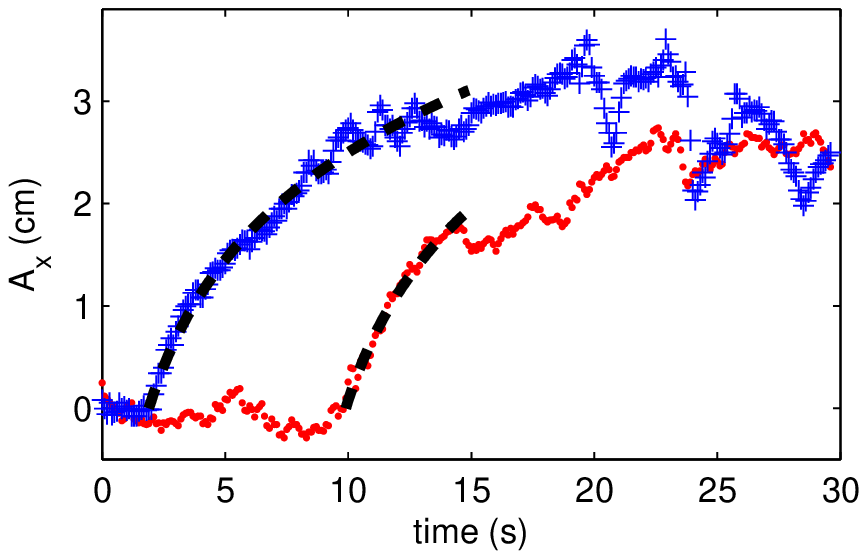}
\caption{Symmetry breaking data for strong symmetry breaking, depth (top) and asymmetry (bottom): this figure shows two experiments (plotted in red and blue) with identical initial conditions ($P=6$~psi, $h=35$~mm). The symmetry breaking occurs at different times for the two different experiments. The symmetry breaking shows no preferential behavior as far as when it will occur or which 
side it will move toward.}
\label{strongdata}
\end{figure}

\begin{figure}[th]
\centering
\includegraphics[width=0.9\columnwidth] {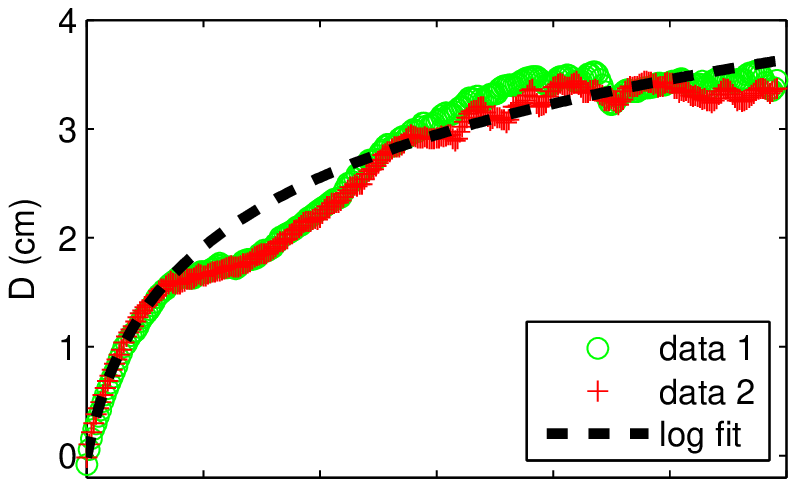}
\includegraphics[trim=0mm 0mm 0mm 6mm, width=0.9\columnwidth] {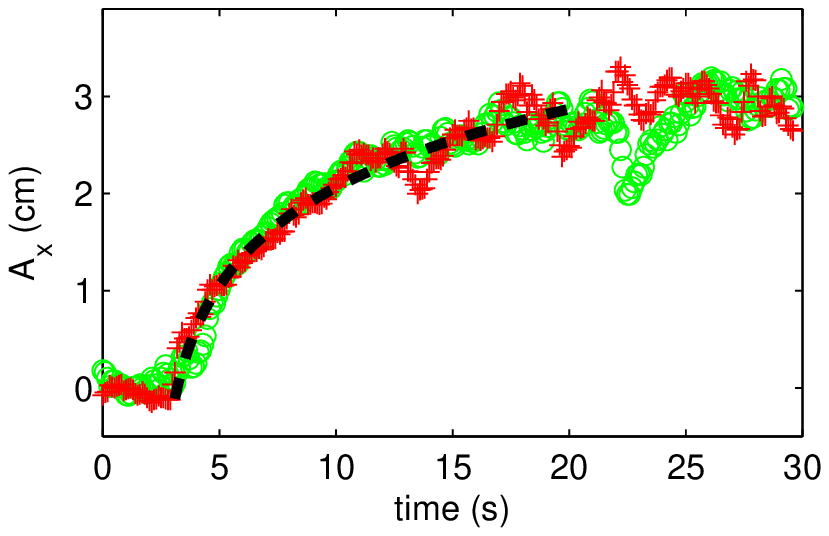}
\caption{Data for strong symmetry breaking, depth (top) and asymmetry (bottom): this figure shows two experiments (plotted in red and green) with identical initial conditions ($P=7$~psi, $h=35$~mm) such that symmetry breaking occurs at the same time after initiation of the air flow. Note the similarity between the two experiments for both the depth, $D$ and asymmetry, $A_x$.}
\label{strongdata2}
\end{figure}

In contrast, for the strong symmetry-breaking case, the logarithmic growth of the depth is severely disrupted when the symmetry is broken, and the horizontal position changes much more rapidly. Fig.~\ref{strongdata} shows the depth and horizontal positions of craters for two separate experiments for the same initial conditions. Both data sets start along the same path, where $D(t)$ follows a logarithmic function and $A_x$ is zero. The red data follows this trajectory much longer than the blue data. Each set spontaneously breaks from the logarithmic function at a different time. At this time, the depth growth is at least temporarily interrupted and the asymmetry, $A_x$, begins to grow, following a logarithmic evolution curve, relatively smoothly at first, and then with fluctuations imposed. During the latter stages of evolution, the depths for the two different cases approach each other, and likewise the horizontal positions of the two data sets approach each other. In fact, the rather complex structures for depth and asymmetry can be reproducible if the symmetry breaking occurs at the same time after initiating the flow. For instance, Fig.~\ref{strongdata2} shows two experiments with the same initial conditions which happen to break symmetry at the same time. Note the consistency in the behavior, both in the symmetric growth period (logarithmic depth growth) and in the symmetry breaking period (quasi-logarithmic horizontal change).

There is an additional insight that can be made by observing different types of digging and the corresponding crater geometry. As discussed previously, a larger initial height of the nozzle yields a broader crater, and a lower nozzle yields a narrower crater. Most importantly, as mentioned previously, material leaves more vertically from narrower craters than for broad craters. The direction in which granular material is ejected has the effect of essentially coupling or decoupling the debris leaving the two sides of the crater and then returning. For broad craters, it is extremely unlikely that debris leaving one side of the crater ends up on the other side. However, for very narrow craters, a significant portion of the debris exiting one side can be directed to the opposite side, which can strongly affect horizontal asymmetries. That is, if there is a narrow crater where the coupling between the two sides is extremely strong, once the digging becomes slightly more prominent on one side, a substantial portion of that debris is dumped back into the other side. This effect appears to drive the strong symmetry-breaking case. 

\begin{figure}[h!tbp]
\centering
\includegraphics[width=0.9\columnwidth]{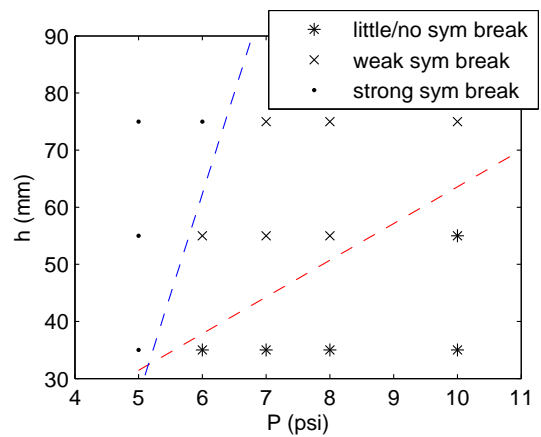}
\caption{Plot shows where different kinds of symmetry breaking tend to occur as a function of the starting nozzle height ($h$) and the driving pressure for the gas ($P$). Dashed lines are meant to suggest different regions where each type of cratering occurs, but further investigation would be required to adequately map out these regions.}
\label{symbreakdata}
\end{figure}

We identify strong symmetry-breaking as the case where the sidewalls of the inner crater range from nearly vertical to overhanging and a circular-like flow pattern emerges: the flow tends to go down one edge of the crater and upward and out the other, throwing debris back into the downward flow and into the other side of the crater. Likewise, we define weak symmetry-breaking as the case where the sidewalls are not as steep and there is not a circular-like pattern, but where there is a clear preferential direction to the digging, albeit not as severe. With these definitions in mind, Fig. \ref{symbreakdata} indicates when each type of symmetry-breaking occurs as a function of the starting height of the nozzle and the driving pressure. In summary, symmetry breaking does not occur for very large $h$ and/or small $P$. When it occurs, smaller $h$ is necessary for a given $P$ if strong symmetry breaking is to occur. 

This symmetry breaking is strongly suggestive of a pitchfork bifurcation associated with the the asymmetry, $A_x$, in the horizontal direction. In such a bifurcation, an increase in the relevant control parameter causes a single stable fixed point to become unstable to two stable symmetric fixed points. Such a bifurcation could be either forward or backward, where the latter is characterized by hysteresis and a discontinuous jump in the steady state value of $A_x$ as the control parameter is raised above the transition. 

First we look at the continuity of the of $A_x$ during symmetry breaking. In the strong symmetry breaking case, it appears discontinuous (see Figs. \ref{strongdata} and \ref{strongdata2}), while in the weak symmetry breaking case, it seems much slower and less abrupt (see Fig. \ref{weakdata}). This suggests that perhaps the strong symmetry breaking is backward and the weak symmetry breaking is forward.

\begin{figure}[h!tbp]
\centering
\includegraphics[width=\columnwidth]{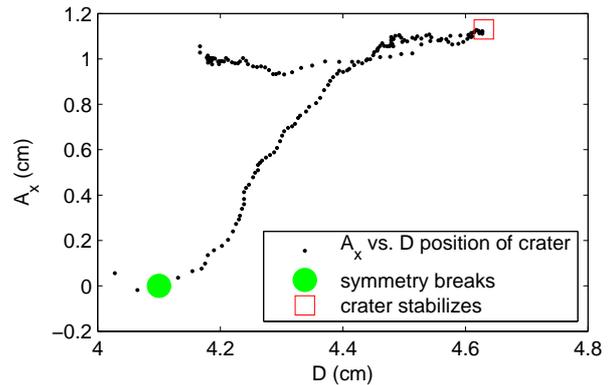}
\caption{Characteristic data for an ``adiabatic'' process (see text) with a jet height $h=35$~mm, where the pressure is varied sufficiently slowly between 3 and 6~psi, such that the crater evolution follows a succession of nearly steady states. The data set begins just before the onset of symmetry breaking. At 6~psi, the crater breaks symmetry and runs to one side, as the crater deepens slightly. Video shows this to be a strong symmetry breaking transition, and the system parameters ($P=6$~psi, $h=35$~mm) are consistent with the strong regime in Fig.~\ref{symbreakdata}. After the crater has stabilized, the pressure is slowly decreased to a value which is below the initial onset of symmetry breaking. The crater remains fully asymmetric during this process, showing no signs of returning to the middle.}
\label{hysteresis}
\end{figure}

We can also look for hysteresis to confirm these classifications. By operating the system in an approximately adiabatic mode, where the pressure is varied slowly in small increments so the depth stays approximately at its asymptotic value, one can clearly observe hysteresis at a jet height (and crater shape) characteristic of strong symmetry breaking processes. Figure~\ref{hysteresis} shows data from an example of this procedure, with a jet height of 35 mm. As the pressure is slowly increased, the crater grows deeper. At some point, the system breaks symmetry. Once the symmetry breaking occurs, the pressure is held constant until the system stabilizes. Then, the pressure is slowly decreased to a level well below the value for the original symmetry breaking. However, the amount of asymmetry remains effectively constant. 

For weak symmetry breaking (i.e. with a jet height of 75 mm and a wide crater shape), no sudden symmetry breaking occurs in an adiabatic mode of operation. Instead, the crater might shift a very small amount, but it is always possible to reduce the pressure again and have the system remain stable and relatively symmetric. Note that even a symmetric cratering process is innately hysteretic and irreversible (i.e., if a significant crater is created, the system can never return to a crater-less state simply by changing the gas flow in some way, such as varying the pressure or jet height). In this way, hysteresis exists in both types of symmetry breaking processes.

\section{Discussion and Conclusions}

\subsection{Analysis of Logarithmic Growth}

Given the ubiquity of a logarithmic form for cratering from fluid flows,  
it is useful to consider mathematical models, i.e. one or more differential equations, which would give rise to such a form. For instance cratering experiments including those of Metzger et al., the present experiments, and even experiments with a disturbance coming from below a submerged granular bed \cite{varas}, modeling a volcanic eruption, exhibit logarithmic growth. 
At the simplest level, one might write the following evolution equation for $D$:
\begin{equation}
\frac{dD}{dt}=a b \exp{(-D/a)}.
\label{depthexp}
\end{equation}
This equation contains both a length scale, $a$, and a time scale, $b$. The length scale may well be tied to the spreading of the gas plume as it approaches the granular bed (see discussion above). The gas speed (determined by the pressure driving the flow) and $a$ are then sufficient to determine a time scale. In fact, a somewhat different form for the evolution
of $D$ is also useful. Defining $u = \dot{D}$, yields an evolution equation
\begin{equation}
\frac{du}{dt} = -u^2/a
\label{ueq}
\end{equation} 
which does not contain an explicit time scale. Integrating to obtain $u(t)$, yields
\begin{equation}
u = \frac {dD}{dt} = u_o a/(a + u_o t)
\end{equation}
where $u_o$ is the initial value of $dD/dt$, which has units of velocity. Identifying $u_o$ with the gas jet velocity seems reasonable, in the absence of additional information. The logarithmic form for $D(t)$ then follows as
\begin{equation}
D(t)= a \ln[1 + (u_o t/a)].
\label{newlogform}
\end{equation}
We note that the cutoff in the logarithmic growth can easily be modeled by:
\begin{equation}
\frac{dD}{dt}=a b \left[\exp{(-D/a)}-\exp{(-D_\infty /a)}\right],
\label{modifieddiffeq}
\end{equation}
where $D_\infty$ is the saturation depth. This equation leads to the following time dependence:
\begin{equation}
D(t) =a \ln{\left(e^{\frac{D_\infty }{a}}-(e^\frac{D_\infty }{a}-1)e^{-bt \exp {(-D_\infty /a)}}\right)}.
\label{modifieddepth}
\end{equation}
For short times, we recover the previous form:
\begin{equation}
D(t) \approx a \ln{(bt+1)}.
\end{equation}
For long times, the system approaches an asymptotic value of
\begin{equation}
D(t\rightarrow\infty) =D_\infty .
\end{equation}

Revisiting the data in Fig.~\ref{saturation}, the functional form of Eq.~\ref{modifieddepth} is shown to be an acceptable fit to the crater depth in the logarithmic and saturation regimes. These results are indicated by the thin red line.

\subsection{Conclusions}

The present experiments have explored the nature of cratering caused by impinging gas jets on quasi-2D samples of various granular materials, including lunar and Martian simulants and terrestrial sand. We observe initially logarithmic grow of the crater depth. We also observe what we believe is a previously unreported symmetry breaking instability. Assuming that a similar effect occurs in 3D, this symmetry breaking could be of extreme practical concern for the design of a spacecraft which needs to land on or take off from a granular surface. A lander design based on symmetric crater evolution could lead to unexpected catastrophic results if a symmetry breaking instability occurs and the crater moves laterally, disrupts the surface where the lander should make contact with the ground, or if the exhaust from the rocket, along with entrained material, is redirected towards the lander. For instance, an imbalance or redirection of the thrust or exhaust stream might exert enough torque to flip it over. We also note that previous work \cite{metzger6,metzger7} has suggested that planetary landings, especially on less frictional granular surfaces (i.e., Mars), may have other potential failure modes.

An interesting analogue may occur in experiments involving water droplets hitting a bed of sand covered by a thin layer of water. In the experiments of Kellay \cite{kellay} droplets created a circularly symmetric disturbance for a 3D sand bed, and at a certain drop frequency, the circular symmetry was broken, leading to a logarithmically evolving asymmetric pattern. 

\begin{acknowledgements}

This work was supported by ORBITEC (contract \# OTC-GS-02381), subcontracted from USAF (contract \# NNX09 CF72P) 
and by NASA contract. We particularly thank Dr. Phil Metzger for input on various aspects of the problem of crater formation. 

\end{acknowledgements}



\begin{thebibliography}{10}

\bibitem{metzger1} P.T. Metzger, R.C. Latta, J.M. Schuler, C.D. Immer, AIP Conference Proceedings 7/1/2009, Vol. 1145 Issue 1, 767-770

\bibitem{metzger2} P.T. Metzger, et al, Journal of Aerospace Engineering Jan 2009, Vol. 22 Issue 1, 24-32

\bibitem{metzger4} Metzger, P., Lane, J., Immer, C., Gamsky, J., Hauslein, W., Li, X., Latta, III, R., and Donaue, C. Earth and Space 2010: pp. 191-207, (2010)

\bibitem{metzger5} LaMarche, C., Curtis, J., and Metzger, P. Earth and Space 2012: pp. 36-44 (2012)

\bibitem{metzger3} K.J. Berger, Anshu Anand, P.T. Metzger, and C.M. Hrenya1, Phys. Rev. E \textbf{87}, 022205 (2013)

\bibitem{Kuang} S.B. Kuanga, C.Q. LaMarcheb, J.S. Curtisb, A.B. Yua. Powder Technology \textbf{239} May 2013, pp. 319-336

\bibitem{Kok} Jasper F. Kok, Eric J. R. Parteli, Timothy I. Michaels, and Diana Bou Karam Rep. Prog. Phys. \textbf{75} 106901 (2012)

\bibitem{rajaratnam} N. Rajaratnam, J. Hydr. Div. Vol. 103, No. 10, October 1977, 1191-1205

\bibitem{mazurek} K.A. Mazurek, N. Rajaratnam, and D.C. Sego, J. Hydr. Engrg. Volume 127, Issue 7, 598-606 (July 2001)

\bibitem{varas} G. Varas, V. Vidal, and J-C Geminard, Phys. Rev. E \textbf{79}, 021301 (2009)

\bibitem{mckay} D.S. McKay, et al, Engineering, Construction, and Operations in Space IV, American Society of Civil Engineers, 857-866, 1994

\bibitem{allen} C. C. Allen, R. V. Morris, D. J. Lindstrom, M. M. Lindstrom, and J. P. Lockwood. JSC Mars-1: Martian regolith simulant, Lunar Planet. Sci., 282, abstract 1797 (1997)

\bibitem{metzger6} P.~T. Metzger, R.~P. Mueller, \& P.~E. Hintze. LPI Contributions, 1679, 4359 (2012)

\bibitem{metzger7} Philip Metzger, Xiaoyi Li, Christopher Immer, John Lane. 47th AIAA Aerospace Sciences Meeting, January 5-8, 2009. doi:10.2514/6.2009-120

\bibitem{kellay} H. Kellay, Europhys. Lett., \textbf{71} (3), pp. 400-406 (2005)

\end{thebibliography}
\end{document}